\documentclass{aa}  
\usepackage{graphicx}
\usepackage{subfigure}
\usepackage{txfonts}
\usepackage{xcolor}
\usepackage{mhchem}     
\usepackage{amsmath}
\newcommand{\rev}[1]{#1} 
\usepackage{hyperref}
\begin{document}

   \title{Ion-molecule
   routes towards cycles in TMC-1}
   \subtitle{An automated study of the \ce{C2H4 + CH2CCH+} reaction }

   \author{M. Mallo,
         \inst{1}
          M. Ag\'undez,
          \inst{1}
          C. Cabezas,
          \inst{1}
          O. Roncero,
          \inst{1}
          J. Cernicharo,
          \inst{1}
          G. Molpeceres,
          \inst{1} 
          }
   \authorrunning{Mallo et al}
   \institute{Departamento de Astrofísica Molecular, Instituto de Física Fundamental (IFF-CSIC), C/ Serrano 123, 28006 Madrid, Spain\\
      \email{maria.mallo@iff.csic.es; german.molpeceres@iff.csic.es}
             }

   \date{Received \today; accepted \today}

 
  \abstract
{Cyclopentadiene (\ce{c-C5H6}) is considered a key molecule in the formation of polycyclic aromatic hydrocarbons (PAHs) in the insterstellar medium (ISM). The synthesis of PAHs from simpler precursors is known as the "bottom-up" theory, which, so far, has been dominated by reactions between organic radicals. However, this mechanism struggles to account for the origin of the smallest cycles themselves. Ion-molecule reactions emerge as promising alternative pathways to explain the formation of these molecules.
  
In the present work, we investigate the reaction network of the main ionic precursor of cyclopentadiene, \ce{c-C5H7+}. To this end, we establish an integrated protocol that combines astrochemical modelling to identify viable formation routes under cold interstellar medium conditions, automated reaction path search and kinetic simulations to obtain accurate descriptions of the reaction pathways and reliable rate constants. In particular, we examine the reaction between ethylene (\ce{C2H4}) and the linear propargyl cation (\ce{CH2CCH+}). Our results reveal that the formation of \ce{c-C5H7+} by radiative association turns out to be inefficient, contrary to our initial expectations. Instead, the system predominantly evolves through bimolecular channels yielding \ce{c-C5H5+} and \ce{CH3CCH2+}, with the formation of \ce{c-C5H5+} offering new insights into reactivity that supports molecular growth in the ISM.}



      \maketitle
%
\newcommand{\kcalmol}{kcal mol$^{-1}$ }
\section{Introduction}

Polycyclic aromatic hydrocarbons (PAHs) emerged as plausible carriers of the unidentified infrared emission (UIR) bands in the early 1980s \citep{leger, allamandola1985}, being hypothesized then to be present in UV-irradiated interstellar regions. The compatibility with these set of bands, summed to the unique properties of PAHs as, for example, charge regulators or interstellar catalysts make them arguably most important family of interstellar molecules \citep{tielens_physics_2005}. The recent detection of a large number of PAHs in the Taurus Molecular Cloud (TMC-1) \citep{cernicharo2021pure, mcguire2021detection, acenaphtylene, cyanopyrene, cyanopyrene2, cyanocoronene, cabezas_discovery_2025} has shaken up the astrochemical community, which still discusses the implications of the large reservoir of PAHs found and opens a new window into an understanding of the chemistry of complex molecules in the interstellar medium (ISM). 
The chemical processes leading to PAHs formation is a widely debated subject nowadays \citep{kaiser2003elementary, joblin, thomas2019barrierless, concepcion, goettl2025gas} and still lacks a satisfactory explanation in such cold and UV-shielded environments. Among the proposed mechanisms, the bottom-up scenario for molecular mass growth supports that small hydrocarbons are first synthesized and then grow into larger structures \rev{\citep{agundez2025search, goettl2025molecular, goettl2025gas, castineira2024comprehensive}}. 

Simple molecules such as benzene (\ce{c-C6H6}) and cyclopentadiene (\ce{c-C5H6}) are likely building blocks in the formation of larger aromatic hydrocarbon cycles. However, the formation of these first rings in the interstellar medium is not yet fully understood. Therefore, an exhaustive search for new routes that could lead to the formation of these molecules is required.  While typically neutral-neutral reactions have been invoked in the formation of interstellar cycles, \rev{\citep{jones_formation_2011, Doddipatla2021,concepcion, yang_low-temperature_2024, castineira2024comprehensive}}, these processes alone cannot account for the large abundances found for interstellar PAHs. 
In this context, ion-molecule reactivity, on top of which most of the complexity build-up in cold environments is based upon constitutes a type of processes candidate to explain the synthesis of aromatic cycles.

In molecular databases, the formation of cyclopentadiene is proposed to proceed through the dissociative recombination of its ionic precursor, the cyclopentadienyl cation (\ce{c-C5H7+}), with electrons:
\begin{equation}
   \ce{c-C5H7+ + e- -> c-C5H6 + H} \label{eq:reaction}
\end{equation}
which represents a common pathway to increasingly complex carbon bearing molecules, although the branching ratios for the possible neutral products are usually undetermined \citep{herbst2021}. This mechanism is analogous to the one proposed to synthesize benzene \citep{mcewan, woods2002synthesis}, whose presence in cold dense clouds is inferred from the detection of the cyano derivative \citep{mcguire2018detection, cooke2020benzonitrile}. We note that route to the formation of benzene has recently been questioned \citep{kocheril2025termination} although the scientific debate is still opened at the time of writing this article \citep{loison2025evidence}.

Cyclopentadiene was recently detected in the cold dark cloud TMC-1, alongside ethynyl cyclopropenylidene (\ce{c-C3HCCH}) and indene (\ce{c-C9H8}) by \cite{cernicharo2021pure}.
However, the mechanisms underlying its formation from non-cyclic hydrocarbon precursors, which would support theories of molecular mass growth in interstellar environments, still remain under discussion \citep{cyclopentadiene, zhang2025formation}
and current chemical models are unable to explain its observed abundance \citep{mcguire,fulvenallene}.

 
Returning to \ce{c-C5H7+}, and according to the Anicich review on bimolecular gas phase cation-molecule reaction kinetics \citep{anicich}, the reaction between the propargyl cation (\ce{CH2CCH+}) and ethylene (\ce{C2H4}) is one of the few reactions that has been experimentally studied and reported to produce \ce{c-C5H7+} \citep{smyth1982ion}. Incorporating this reaction into chemical networks has been demonstrated to improve the description of \ce{c-C5H6} in TMC-1 \citep{fulvenallene}. Said reaction also leads to \ce{C5H5+ + H2}, but the branching ratio for the products is unknown. Ethylene has been detected in circumstellar gas surrounding the carbon-rich star IRC+10216 \citep{betz1981ethylene, fonfria2017abundance} and its cyano derivative vinyl cyanide (\ce{CH2CHCN}) was first observed in TMC-1 by \cite{matthews1983detection} and towards IRC+10216 \citep{agundez2008detection}, among several other sources. On the other hand, the propargyl cation presents two isomeric forms: the linear one (\ce{CH2CCH+}) and the cyclic (\ce{c-C3H3+}). In the present work, we focus on \ce{CH2CCH+}, which was recently discovered in TMC-1 \citep{silva2023discovery}. The cyclic isomer, which is more stable and less reactive, has not been detected in space yet, as it does hold a permanent dipole moment. 

Our theoretical study focuses not only on veryfing the formation of \ce{c-C5H7+} but also on determining how competitive are the different channels of the reaction. The study of ion-molecule reactivity is often more complex than that of neutral-neutral reactions, which typically proceed through well-defined pathways such as the loss of \ce{H} or \ce{CH3} \citep{heitkamper2022reactivity, balucani1999crossed, balucani2018theoretical}. On the other hand, ion-molecule processes are not as straightforward as neutral-neutral ones, and the atomic rearrangement is difficult to predict with exactitude. To approach this problem, we employ an automated reaction discovery method that allows us to explore the potential energy surface (PES) of the \ce{C2H4 + CH2CCH+} system in an unbiased manner, identifying all the possible reaction pathways and products within an energy threshold. Our aim is to demonstrate the potential of an automated protocol to explore the reactivity of ion-molecule systems, reaching trustworthy results that can be directly applied to astrochemical models. In this work, we automatically build the reaction network of the \ce{c-C5H7+} molecular formula. We begin by an inspection of potentially relevant ion-neutral reactions leading to \ce{c-C5H7+} aided by chemical models, followed by quantum chemical calculations and automated reaction path exploration techniques to identify the viable reaction pathways.


\section{Methodology} \label{sec:methodology}

\begin{figure*}
   \centering
   \includegraphics[width=\linewidth]{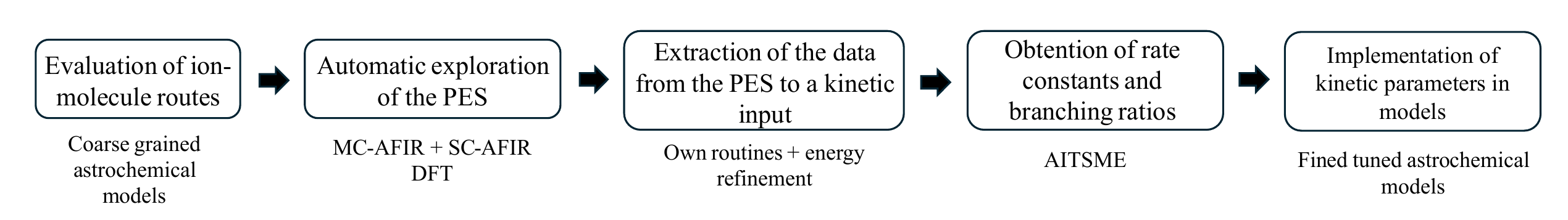}
      \caption{Schematic diagram of our automated protocol to explore the reaction network of ion-molecule systems.}
         \label{fig:diagram}
\end{figure*}

Exploring the potential energy surface (PES) of complex ion-molecule reactions is a challenging task where manual searches are highly prone to error. To overcome this, we employ an unbiased search for stationary points, an area of computational chemistry that has rapidly advanced in recent years \citep{sameera_computational_2016, maeda_artificial_2016, maeda_exploring_2021, van_de_vijver_kinbot_2020, Martinez-Nunez2021, maeda2023toward, turtscher_span_2023} \rev{Such methodologies have been even used to explore astrochemically relevant systems \citep{castineira2024comprehensive, komatsu2022quantum, komatsu2023automated}.}
Among the most successful approaches is the artificial force induced reaction (AFIR) method \citep{maeda_artificial_2016}, which applies artificial pushing and pulling forces between fragments to explore reactivity.

In this work we use the AFIR method as implemented in the \textsc{GRRM23} (Global Reaction Route Mapping) program \citep{maeda_afir_2013, maeda2023toward, grrm23}. AFIR generates plausible reaction pathways by introducing attractive or repulsive forces within a molecule, and whenever a new stationary point on the PES is located it is added to the pool of candidate structures from which the search continues, ultimately building a global (or semi-global, if constraints are applied) reaction network. To avoid excessive computational cost in high-energy regions, all simulations were terminated according to some predefined stopping criterions: (i) an upper energy threshold that corresponds to the energy of the reactants including zero-point energy (ZPE), and (ii) a convergence rule in which the search was stopped when last 30 explored paths did not update the lowest 15 local minima.
The central parameter controlling the amount of energy introduced into the system is the collision parameter $\gamma$, which approximates the largest barrier that can be overcome in the search. To ensure an adequate exploration of stationary points, including bimolecular dissociation channels, we employed a value of $\gamma = 400.0$ kJ mol$^{-1}$ for a single biasing force (AFIR function). This corresponds to the SC-AFIR method, where "SC" (single component) indicates that the search starts from a potential well, in our case the lowest-energy structure formed by the barrierless association of \ce{C2H4} and \ce{CH2CCH+} at long range. This initial association event was simulated separately using the MC-AFIR protocol \citep{Maeda2018}, where "MC" (multicomponent) indicates that the two fragments, \ce{C2H4} and \ce{CH2CCH+}, were placed at long distance and pulled together by a single artificial force of $\gamma = 100$. A total of 100 sampling runs were performed in this stage, and the AFIR profiles confirmed the barrierless insertion of the fragments. All AFIR searches were performed with \textsc{GRRM23}, interfaced with \textsc{Gaussian16} \citep{g16} for electronic structure calculations.

The SC-AFIR calculations are performed at the $\omega$B97XD/6-31+G* level \citep{omega, hehre1972self, hariharan1973influence}, and, as indicated above, only considering exothermic and barrierless pathways due to the low temperatures of TMC-1.  Refinement calculations are performed using a higher level of theory to characterize the stationary points, optimizing them at the $\omega$B97XD/def2-TZVPD level \citep[][in addition to the original references of the method included above]{def2, rappoport_property-optimized_2010}, from which we also extract the harmonic vibrational frequencies and zero-point energy (ZPE) corrections. 
\footnote{The def2-TZVPD basis set is included in \textsc{Gaussian16} from the corresponding entry of the Basis Set Exchange Library \url{https://www.basissetexchange.org/} \citep{pritchard_bse_2019}}
Finally, the electronic energies are corrected at the DLPNO-CCSD(T)/aug-cc-pVTZ level \citep{purvis_full_1982,dlpno,guo_communication_2018,aug-cc2, papajak2011perspectives} using a tight localization scheme (\texttt{tightpno} in \textsc{Orca} settings). Therefore, the method employed for all the refined stationary points presented in this work can be abbreviated as DLPNO-CCSD(T)/aug-cc-pVTZ//$\omega$B97XD/def2-TZVPD.

Once the potential energy surface (PES) is explored and characterized, a kinetic investigation is performed to determine the rate constants for the different pathways. The theoretical approach to get the global rate constants relies on the ab-initio transition state based master equation (AITSME) under a microcanonical formalism. To build the master equation, the Rice-Ramsperger-Kassel-Marcus (RRKM) theory is used to calculate the unimolecular elementary steps of the reaction. Furthermore, the barrierless channels are modeled using the phase-space theory  \citep{Pechukas1965,Truhlar1969}, in which the long-range interaction between the reactants is described by a potential form $V(r)$ of the type:
\begin{equation}
   V(r)=-\frac{C_{4}}{r^{n}},
 \end{equation}
The $V(r)$ points are obtained from rigid potential energy scans at the MP2/aug-cc-pVTZ level of theory \citep{moller1934note} between distances of 4-25 \r{A}. 
The global rate constants are derived from the individual ones using the master equation formalism implemented in the MESS code \citep{mastereq, mess}.

The protocol connecting the automated search of the PES to the kinetic analysis is unsupervised thanks to a combination of in-house \texttt{Python} routines for the analysis, construction and refinement of the PES. In brief, the stationary points of coming from the \textsc{GRRM23} code are parsed, refined at a higher level of theory, with an on-the-fly construction of the \textsc{MESS} input file. The integration between the different stages of our simulation is carried out within our codes using the \textsc{Ash} library for atomistic simulations \citep{bjornsson2022ash} interfaced with the \textsc{Orca} (v6.0.1) code \citep{Neese2020, ORCA6}. Figure \ref{fig:diagram} illustrates the methodology followed in this work, highlighting the integration of the automated reaction path search with the rate constants calculations through our own routines.

Finally, the rate constants for each bimolecular channel are fitted into the Arrhenius-Kooij equation:
\begin{equation} \label{eq:arrhenius}
   k(T)=\alpha \left(\frac{T}{300}\right)^{\beta} \exp{\left(-\frac{\gamma}{T} \right)},
\end{equation}  
where $\alpha$, $\beta$ and $\gamma$ are determined from fitting the calculated rate coefficients over a given temperature range to the expression \ref{eq:arrhenius}. This way, the equation can be easily implemented in chemical models.


\section{Results} \label{sec:results}

\subsection{Evaluating ion–molecule routes to synthesize \ce{c-C5H7+}} \label{sec:models}

As a prior step to carry out the reaction discovery calculations, we use astrochemical models to identify potentially relevant ion-molecule routes that could lead to the 
\ce{c-C5H7+} cation. The time-dependent gas-phase chemical model is built with the typical parameters of cold dense clouds, which consist on a particle density of \ce{H2} of $2 \times 10^{4}$ cm$^{-3}$, a temperature of 10 K, a cosmic-ray ionization rate of $1.3 \times 10^{-17}$ s$^{-1}$ and a visual extinction of 30 mag (further details in \cite{Agundez2013}). We adopted the so-called low-metal elemental abundances, where the abundance of oxygen is decreased to ensure a C/O ratio of 1, which results in a better overall agreement between calculated and observed abundances in the TMC-1 dark cloud \citep{AgundezCernicharo_inprep}. The chemical network is based on the gas-phase RATE22 network from the \textsc{UMIST} database \citep{millar2024umist}.

We conducted a systematic investigation of ionic-neutral reactions leading to the formation of \ce{c-C5H7+} with 
atomic hydrogen (H) elimination, molecular hydrogen (\ce{H2}) elimination, or through radiative association. 
Each route is evaluated in terms of the stability and abundance of the reactants in TMC-1, as well as whether it has been previously examined and documented in the Anicich index of bimolecular gas-phase cation-molecule reactions \citep{anicich}, thereby ensuring the feasibility of the reactions. 

A rate coefficient of $10^{-9}$ cm$^{3}$ s$^{-1}$ is assumed for all the reactions lacking experimental data, but it should be emphasized that this procedure is only applied for the initial identification of the potential ion-molecule reactions leading to \ce{c-C5H7+} and it is not carried to the chemical models shown later in Section \ref{sec:astro_implications}. This value is an upper limit for ion-molecule rate coefficients based on the measurements in \cite{anicich} so that if a reaction is not found to be relevant at this high rate, it can be discarded. If, on the other hand, a significant impact on the abundance of cyclopentadiene is observed, a more detailed study of the reaction will be performed to derive accurate rate coefficients and check the accuracy of the initial guess.

\begin{figure}
   \centering
   \includegraphics[width=0.8\linewidth]{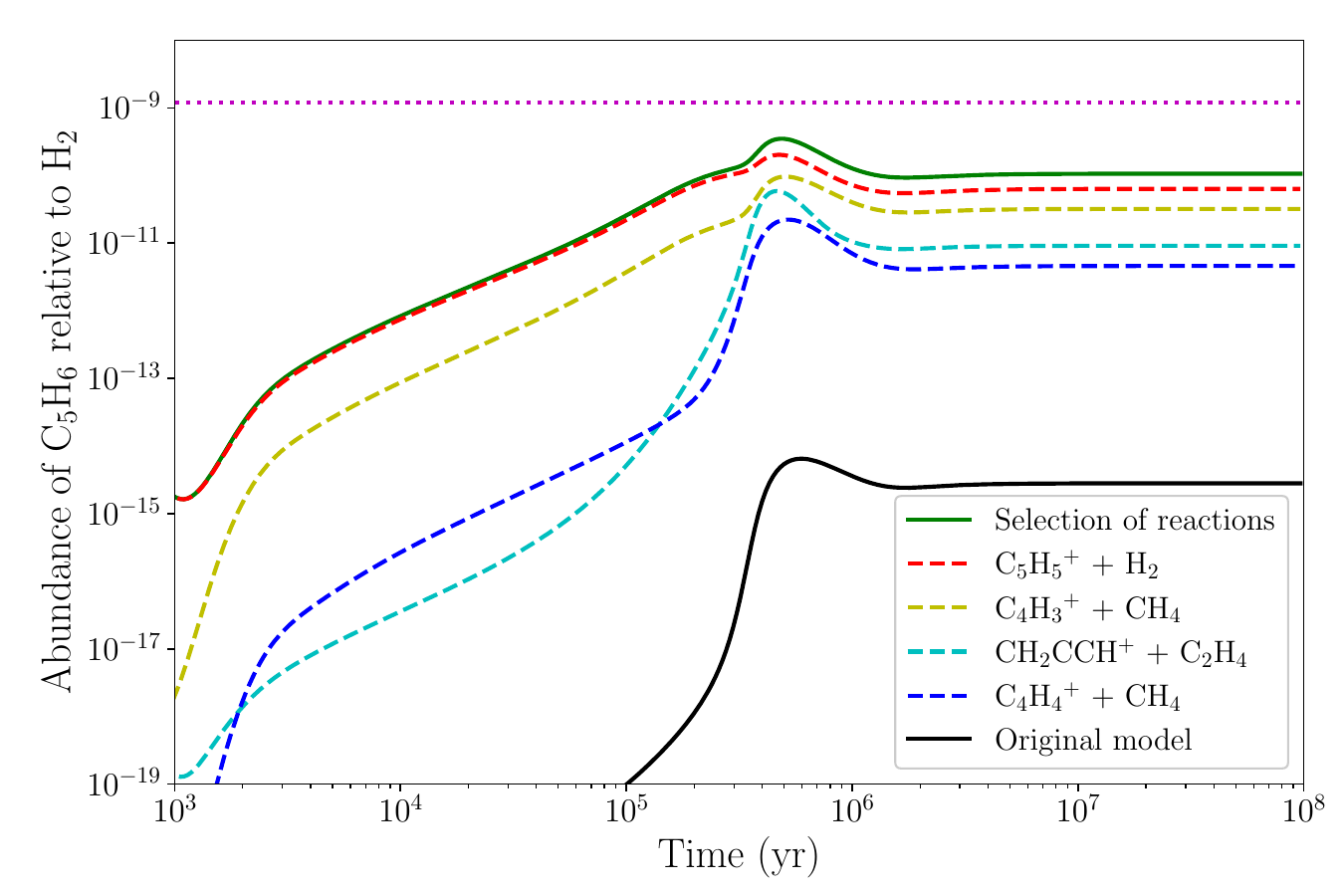}
      \caption{Prediction of the relative abundance of cyclopentadiene (\ce{c-C5H6}) as a function of time in the chemical model of TMC-1. The black solid line is the abundance predicted by the model without ion-molecule reactions, dashed lines represent how it is modified including only each of the reactions separately and the green solid line is the total abundance when all the reactions from the dashed lines are included. The horizontal, dotted line represents the observed abundance of cyclopentadiene in TMC-1, calculated from its column density \citep{cernicharo2021pure}.}
         \label{fig:models}
\end{figure}

With the initial exploration of the reaction network we identified four key reactions that are able to reasonably reproduce the observed abundance of cyclopentadiene within the timescale of $10^{5}$ to $10^{6}$ years according to the chemical model (see Figure \ref{fig:models}). Cylopentadiene is detected with a column density of ($1.2 \pm 0.3$) $\times 10^{13}$ cm$^{-2}$ \citep{cernicharo2021pure}, which results in a relative abundance to \ce{H2} of $1.2 \times 10^{-9}$ adopting a column density of \ce{H2} of $10^{22}$ cm$^{-2}$ for TMC-1 \citep{guelin}. 
The model with the set of reactions that we selected predicts an abundance $4 \times 10^{-10}$ at $10^{5}$ years, which is in close agreement to the observed value. The four reactions that we identified for \ce{c-C5H7+} are the following:

\begin{align}
   \ce{c-C5H5+ + H2 &-> c-C5H7+} + h\nu  \label{eq4} \\ 
   \ce{C2H4 + CH2CCH+ &-> c-C5H7+} + h\nu  \label{eq5}\\  
   \ce{C4H3+ + CH4 &-> c-C5H7+ + h\nu }  \label{eq6}\\ 
   \ce{C4H4+ + CH4 &-> c-C5H7+ + H}  \label{eq7}
\end{align}
After the formation of \ce{c-C5H7+}, it is expected to suffer from electron recombination, i.e., \ce{c-C5H7+ + e- -> c-C5H6 + H}.
The reactions \ref{eq4}-\ref{eq6} represent radiative associations, which consist on stabilization of the product by the emision of a photon. Radiative association is believed to play an important role in the growth of molecules in the cold ISM, where the densities can be too low for collisional stabilization \citep{herbst2021}. Of the four reactions, the reactions involving methane (\ce{CH4}) are not considered in this work due to the presence of energy barriers of 6.1 \kcalmol for the  \ce{C4H3+ + CH4} reaction and 9.0 \kcalmol for the \ce{C4H4 + CH4}, obtained from MC-AFIR explorations. Therefore we turn our attention to the study of reaction \ref{eq5} and, to a lesser extent reaction \ref{eq4}, for which we find kinetic entrance barriers and whose chemistry is investigated in more detail in Section \ref{sec:c5h5_h2}.

\subsection{Automatic exploration of the reaction network} \label{sec:pes}

The first step in the automated exploration of reaction \ref{eq5} consists on identifying the exothermic association complexes between the reactants for each case, as the low temperatures of cold molecular clouds require the existence of open, barrierless entrance channels for the reaction to proceed. 

As previously noted in the Introduction, the propargyl cation in the \ce{C2H4 + C3H3+} system can adopt either a linear (\ce{CH2CCH+}) or a cyclic (\ce{c-C3H3+}) structure. Among these, the cyclic isomer is the most stable form, lying by more than 30 \kcalmol below the linear isomer. Owing to this energy difference between isomers, \ce{c-C3H3+} is expected to be significantly less reactive and not to play a relevant role in this reaction.
We perform preliminary calculations to see the bonding of both isomers with ethylene finding that the \ce{c-C3H3+ + C2H4} association proceeds through activation energies, which can not be overcome at the low temperatures of dense clouds. Then we can discard the cyclic isomer for the further research and only consider the \ce{CH2CCH+ + C2H4} reaction.

\begin{figure}
      \centering
      \subfigure{\includegraphics[width=0.15\textwidth, trim=2cm 4cm 2cm 8cm]{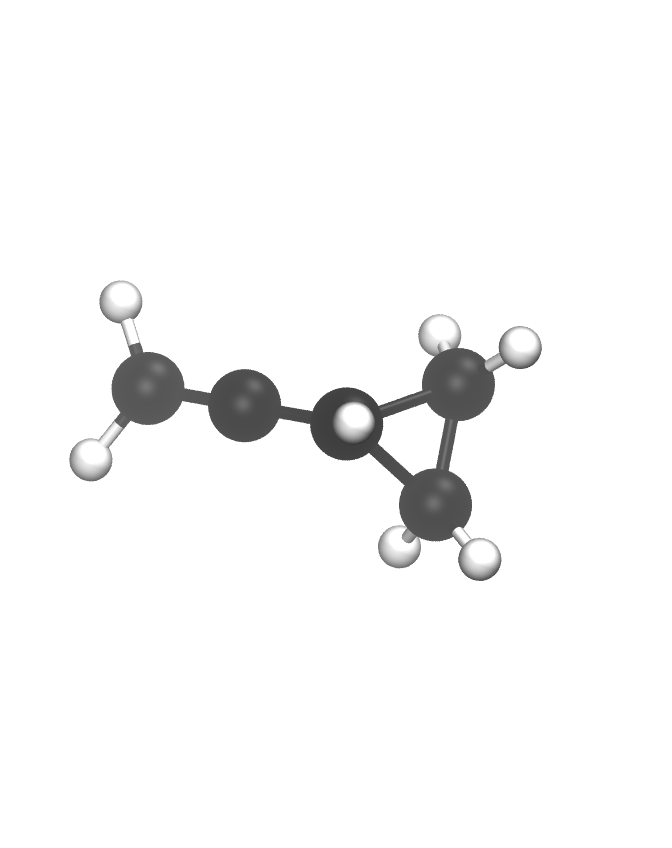}
        \label{fig:r1}}
      \subfigure{\includegraphics[width=0.15\textwidth, trim=0cm 2cm 0cm 5cm]{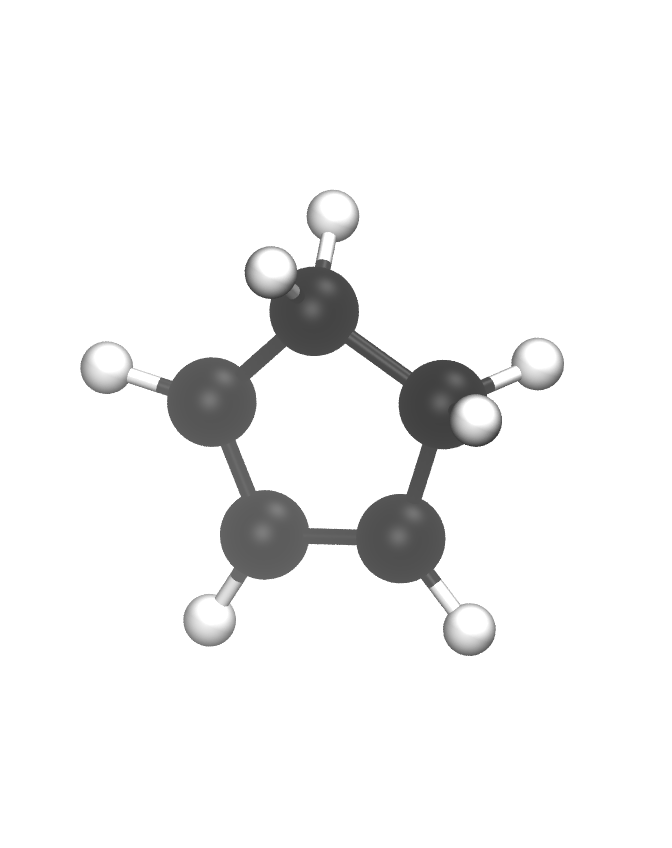}
        \label{fig:r2}}
      \subfigure{\includegraphics[width=0.15\textwidth, trim=0cm 4cm 0cm 6cm]{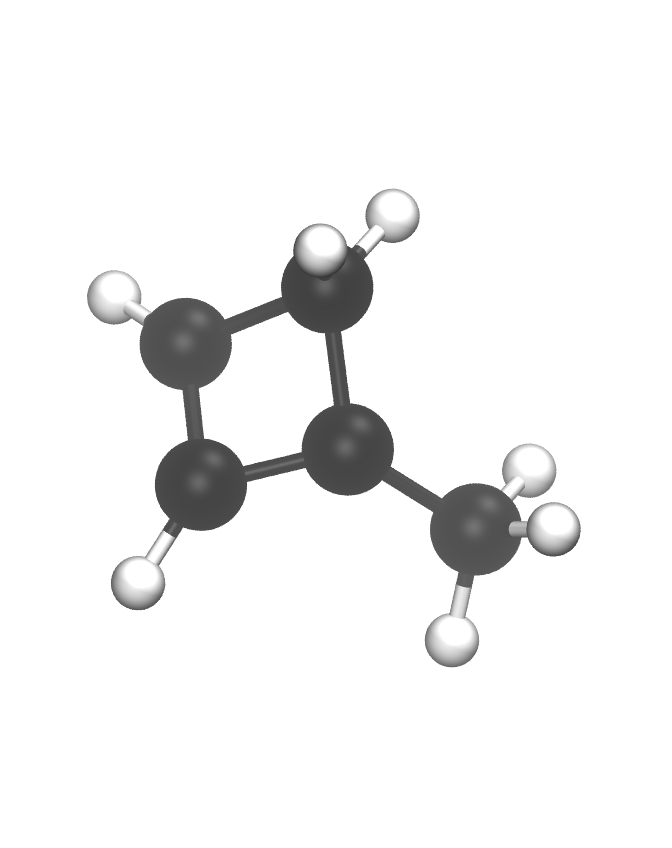}
        \label{fig:r3}}
      \caption{From left to right: structures of R1, R2 and R3 (all of them isomeric forms of \ce{C5H7+})}
      \label{fig:wells}
    \end{figure}

\ce{CH2CCH+} approaches \ce{C2H4} forming the triangular adduct \ce{CH2(CH2)CHCCH2} (refered to in the text as \textbf{R1}, and portrayed in Figure \ref{fig:wells}) with a relative energy of -55.46 \kcalmol. The long-range interaction potential for the \ce{CH2CCH+ + C2H4} association was calculated at the MP2/aug-cc-pVTZ level of theory. The results confirm the absence of energy barriers, in which the capture is characterized by a C$_4$ coefficient of 25.97 au.
According to phase-space theory, the C$_4$ coefficient is related with the Langevin expression for the ion-apolar neutral reaction rate coefficients, predicting that the rates have no dependence on temperature \citep{woon2009quantum}.

\begin{figure*}
   \centering
   \includegraphics[width=\hsize]{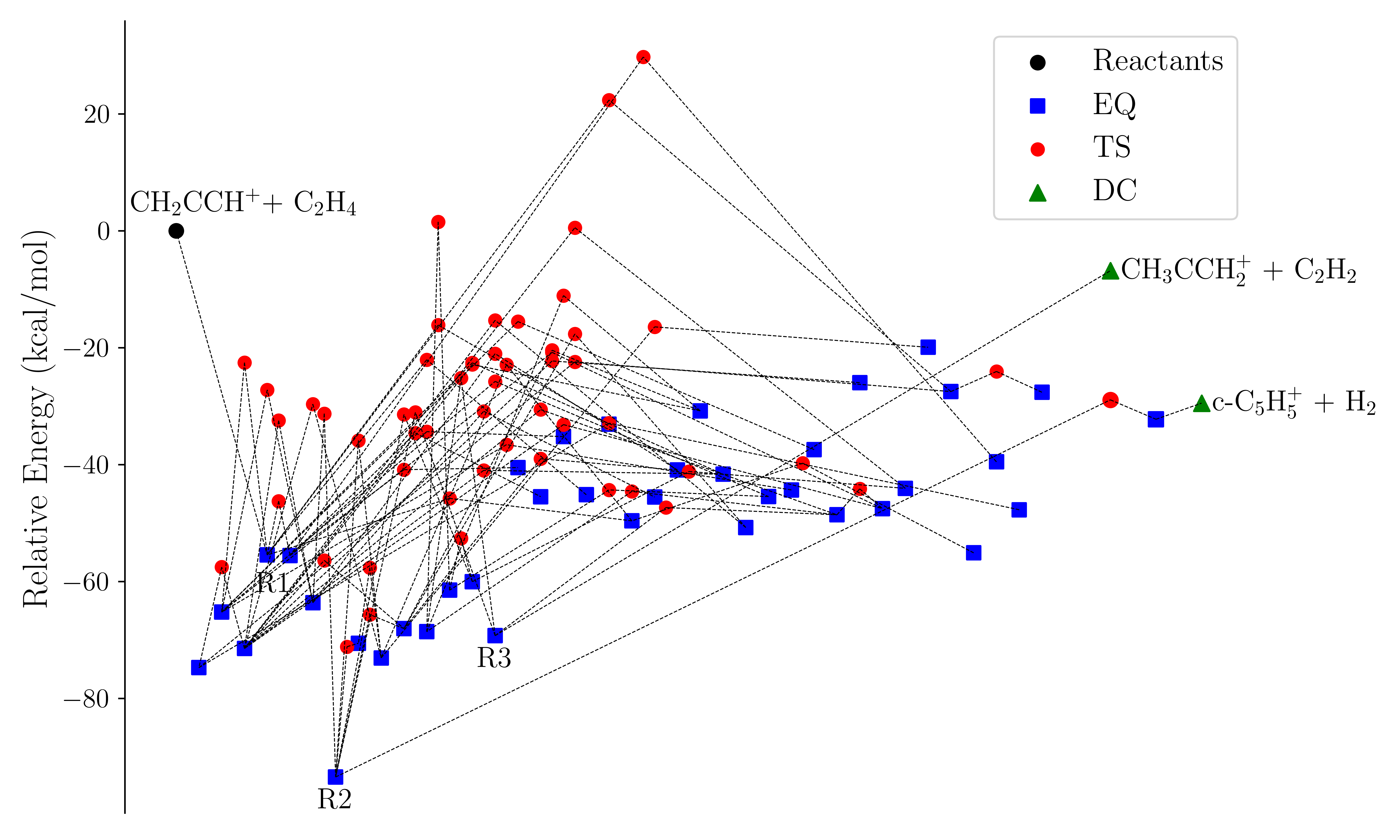}
      \caption{Depiction of the reaction network of the \ce{C2H4 + CH2CCH+} reaction. The energies are calculated at the DLPNO-CCSD(T)/aug-cc-pvtz level with the $\omega$B97XD/def2-TZVPD  harmonic zero point energy. In the diagram EQ are used to refer to the wells (local minima) on the PES, TS refer to transition states, and DC means ``Dissociation channels''. This nomenclature is chosen to be consistent with the one provided by the GRRM code \citep{grrm23}. }
         \label{fig:fullpes}
   \end{figure*}

From \textbf{R1} we perform the automatic protocol described in Section \ref{sec:methodology}. The final energy profile is shown in Figure \ref{fig:fullpes}. The very complex PES landscape evinces that, after the formation of the initial \textbf{R1} adduct, this excess of energy is used by the system to evolve through a series of intermediate wells and transition states among which three paths can be highlighted: the fall into the deepest well \textbf{R2}, which has a relative energy of -93.45 \kcalmol and corresponds to the most stable cyclic isomer of the \ce{c-C5H7+} ion, the dissociation of this well into \ce{c-C5H5+} and molecular hydrogen (\ce{H2}) and the fragmentation from the well \textbf{R3} into the allyl cation (\ce{CH3CCH2+}) and acetylene (\ce{C2H2}). When the system enters the stable well \textbf{R2} it can be stabilized by emission of radiation or evolve to products (\ce{c-C5H5+ + H2}) going through a submerged transition state and a post-reactant complex. The dissociation into \ce{CH3CCH2+ + C2H2} occurs directly from the well \textbf{R3} by sequentially breaking two C-C bonds. To discern which pathways are more favored, kinetic simulations are performed and detailed in Section \ref{sec:kinetic}. However, the presence of barrierless, exothermic bimolecular pathways already indicates that radiative association will be a minor channel compared to bimolecular channels.

\subsection{Rate constant derivation} \label{sec:kinetic}

The kinetic master equation is built considering the stationary points that are found during the automatic exploration of the PES (Figure \ref{fig:fullpes}), all of them being accesible to the system at the low temperatures of TMC-1. 

We focus on the rate constants for the two main, competitive pathways:
\begin{align}
   \ce{CH2CCH+ + C2H4 &-> c-C5H5+ + H2} \label{c5h5} \\ 
   \ce{CH2CCH+ + C2H4 &-> CH3CCH2+ + C2H2}  \label{c3h5}
\end{align}
to which we should add the channel:
\begin{equation}
   \ce{CH2CCH+ + C2H4 -> C5H7+} + h\nu, \label{rad}
\end{equation} 
that represents the stabilization of the complex \textbf{R2} by radiative association.  The presence of very exothermic bimolecular channels greatly reduces the probability of radiative association \citep{herbst2021} in comparison with the two bimolecular ones based in the following competition relation:
\begin{equation} \label{eq:rad_total}
   k_{\rm rad} = \frac{k_{1} \, k_{r}}{k_{-1} + k_{2} + k_{r}}
\end{equation}

where $k_1$ is the rate for the formation of the complex, $k_{-1}$ is the rate for the dissociation of the complex back into reactants, $k_2$ is the dissociation into new products and $k_r$ is the radiative emission rate. Then, the stabilization of the complex by emision of radiation depends on the competition between the re-dissociation and fragmentation into new products. An estimation of $k_{\rm rad}$ based on \citet{tennis_radiative_2021} and \citet{herbst_approach_1982} is derived in Appendix \ref{sec:appA_ratiative_association}, where we determine this channel to be non-competitive with the bimolecular ones with a $k_{\rm rad}$ $\leq$ 3.25$\times$10$^{-14}$cm$^{3}$ s$^{-1}$.

\begin{figure}
   \centering
   \includegraphics[width=0.8\linewidth]{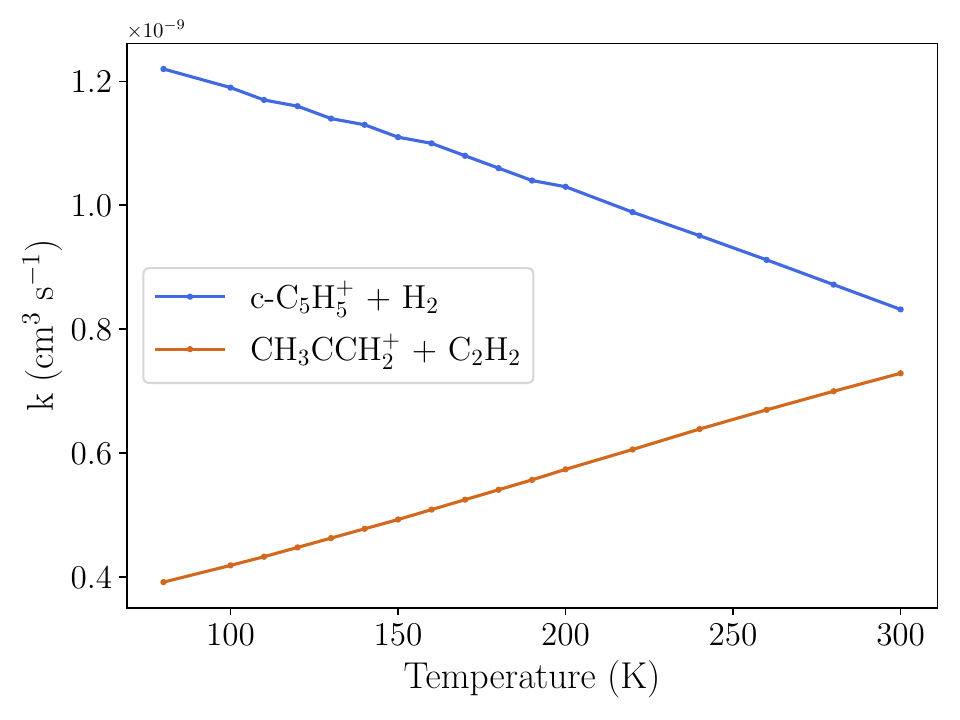}
      \caption{Rate constants (\textit{k}) for the formation of the exothermic products of the \ce{C2H4 + CH2CCH+} reaction as a function of temperature in a range of 80-300 K and a pressure of $10^{-7}$ atm.}
         \label{fig:rates}
   \end{figure}

We compute the rate constants for reactions \ref{c5h5} and \ref{c3h5} in a range of temperatures from 80 to 300 K and a pressure of $10^{-7}$ atm to achieve the most realistic representation possible of the physical conditions of TMC-1. For \ref{c5h5} and \ref{c3h5}, the calculated rate constants are shown in Figure \ref{fig:rates} as a function of temperature. At the lowest temperature achievable in our system using the AITSME approach (80 K), \rev{the rate constant for the formation of \ce{c-C5H5+} is $1.2 \times 10^{-9}$ cm$^{3}$ s$^{-1}$ and the one for the formation of \ce{CH3CCH2+} is $3.9 \times 10^{-10}$ cm$^{3}$ s$^{-1}$.} The rate constant at 80 K are still reasonable estimators of behavior at lower temperature, especially considering that the capture Langevin rate should be temperature independent at low temperatures.\footnote{Only the capture event is strictly temperature independent through the Langevin rate. In contrast, all unimolecular isomerizations within the wells depend on the available energy budget, which includes thermal contributions. At low temperatures this contribution becomes less significant, thereby supporting the validity of the rate constants derived down to 80 K.} The formation of \ce{c-C5H5+} is the dominant channel of the title reaction, \rev{with a branching ratio of 75.7\% versus a 24.3\% of \ce{C3H5+} at 80 K. This ratio shifts to 53.3\% and 47.7\% respectively at 300K}, but the formation of \ce{c-C5H5+} remains the main formation channel. The rate constants of both channels exhibit a significant correlation, in which the formation of \ce{CH3CCH2+} is favored as the formation of \ce{c-C5H5+} is disfavored with temperature. This may be due to the presence of a transition state with a low energy barrier in the complex PES.

\begin{table}
   \centering
   \caption{Arrhenius-Kooij parameters (Equation \ref{eq:arrhenius}) for reactions \ref{c5h5} and \ref{c3h5}. }
   \label{tab:fit}
   \resizebox{\linewidth}{!}{\begin{tabular}{lccc}
   \hline
   Reaction  & $\alpha$ & $\beta$ & $\gamma$\\
   \hline
   \ce{CH2CCH+ + C2H4 -> c-C5H5+ + H2}   & 9.01  $\times 10^{-10}$ & $-0.27$ & $0.00$ \\
   \ce{CH2CCH+ + C2H4 -> CH3CCH2+ + C2H2} & 7.15 $\times 10^{-10}$  & $0.51$  & $0.00$ \\
   \ce{CH2CCH+ + C2H4 -> C5H7+ + h\nu} & 3.25 $\times 10^{-14}$ & 0.00 & 0.00 \\
   \hline
   \end{tabular}}
   \tablefoot{ $\alpha$ in cm$^{3}$s$^{-1}$, $\beta$ is dimensionless and $\gamma$ in K. The derivation of the rate constant for the \ce{CH2CCH+ + C2H4 -> C5H7+ + h\nu} is presented in Appendix \ref{sec:appA_ratiative_association}. \rev{The extrapolation at 10 K leads to uncertainties of a factor of 2-3 in the rate constants.}}
\end{table}
In the experimental study by \cite{smyth1982ion}, the authors report the formation of \ce{C5H7+} and \ce{C5H5+ + H2} with a total rate constant  of $1.1 \times 10^{-9}$ cm$^{3}$ s$^{-1}$ at 300 K. \rev{Our calculations reveal a total rate coefficient of $1.6 \times 10^{-9}$ cm$^{3}$ s$^{-1}$} at the same temperature, which is calculated as sum of the two competitive pathways (k\textsubscript{tot} = k\textsubscript{\ce{c-C5H5+}} + k\textsubscript{\ce{CH3CCH2+}}).
\rev{Despite the excellent agreement with the experimental rate constant of \cite{smyth1982ion}, a significant discrepancy arises from the products of the reaction found}, as they do not report the formation of \ce{CH3CCH2+ + C2H2}, which we find to be a significant channel. The reasons for the disagreement between theory and experiments are not known, and the explanation might be multicausal. A possible source of disparity is the different residence times between experiments and theory. The total absence of the \ce{C2H2 + CH3CCH2+} channels prompt us to consider that secondary reactions involving these molecules might be occurring in the experimental setup, enlarging the total rate constant and including the formation of \ce{c-C5H7+}. This is in contrast with our calculations that are equivalent to single collision experiments, and also to our results in Appendix \ref{sec:appA_ratiative_association} where we use a reduced model to determine the rate constant for radiative association \rev{finding it to be a channel about 4-5 orders of magnitude less eficcient than bimolecular reaction.} However, we encourage contemporary experimental studies. Finally, the \ce{C2H4 + C3H3+} reaction is included in the \textsc{UMIST} database \citep{millar2024umist} with a total rate constant of $1.1 \times 10^{-9}$ cm$^{3}$ s$^{-1}$ leading exclusively to \ce{C5H5+ + H2}. This rate constant can be updated with our results, also including the formation of \ce{CH3CCH2+ + C2H2} with the Arrhenius-Kooij parameters shown in Table \ref{tab:fit}.


\subsection{\ce{c-C5H5+} as product of the reaction and key to new reactivity} \label{sec:c5h5}

\begin{figure}
   \centering
   \includegraphics[width=0.3\linewidth, trim=3cm 3cm 3cm 2cm]{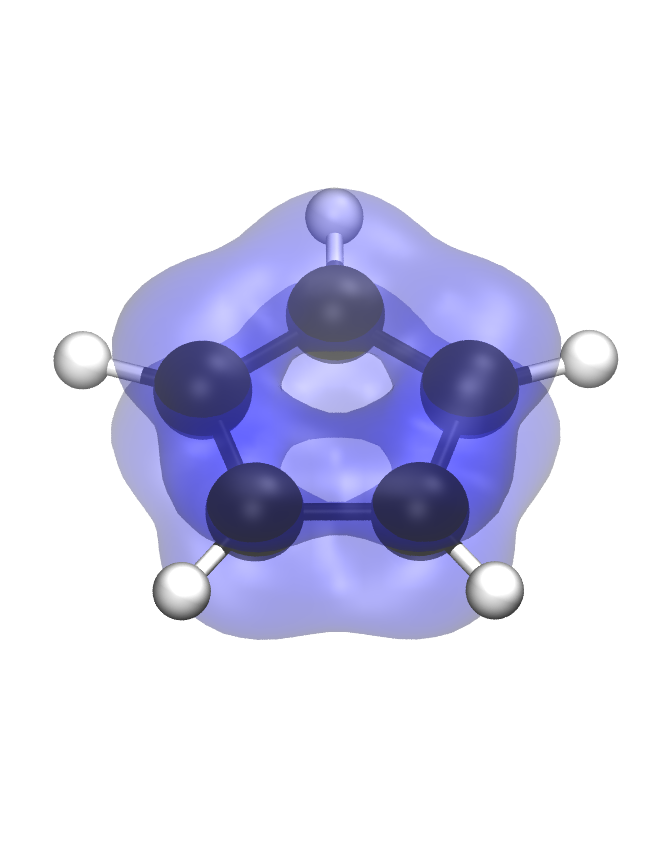}
      \caption{Representation of the spin density distribution of the cyclopentadienylium cation (\ce{c-C5H5+}) in its triplet ground state, calculated at the $\omega$B97XD/def2-TZVPD level of theory. The isovalue is set to 0.005 a.u.}
         \label{fig:density}
\end{figure}

The formation of the cyclopentadienylium cation (\ce{c-C5H5+}) as main product of the \ce{C2H4 + CH2CCH+} reaction may provide a new starting point for the molecular growth in TMC-1. The molecule is a triplet in its electronic ground state, and although the PES shown in Figure \ref{fig:fullpes} is singlet, we assume that it radiatively relaxes after formation.  Its antiaromatic nature makes it a very unstable and reactive molecule, being a promising candidate to continue exploring the bottom-up formation of larger aromatic compounds. Figure \ref{fig:density} shows its corresponding spin density in the ground state, which is evenly distributed across all carbon atoms, suggesting that these sites are equally reactive.

For example, its neutral analog, the cyclopentadienyl radical (\ce{c-C5H5}) has already been considered as potential candidate for the bottom-up formation of PAHs in the recent works \cite{kaiser2021}, \cite{concepcion}. To study the reactivity of \ce{c-C5H5+} and verify its contribution to the interstellar chemistry, we first identified the possible products of the dissociative recombination of \ce{c-C5H5+} with electrons by performing a SC-AFIR exploration of the local and global minima along the PES of \ce{c-C5H4}, adopting a value of $\gamma = 1000.0$ kJ mol$^{-1}$. Initial explorations were carried out using the semiempirical PM7 method \citep{stewart2013optimization}, which provides an optimal balance between computational efficiency and accuracy for large molecular systems. The most stable structures identified in this initial phase were subsequently refined at the $\omega$B97XD/def2-TZVPD level of theory to ensure the reliability of the results, with the energies being calculated at the DLPNO-CCSD(T)/aug-cc-pVTZ level with the ZPE correction.
The prediction of the electron recombination  products is highly uncertain, this is why we assume that the reaction mainly procceeds through hydrogen elimination:

\begin{equation}
   \ce{c-C5H5+ + e- -> C5H4 + H} \label{eq:c5h4}
\end{equation}

\begin{figure}
   \centering
   \includegraphics[width=\linewidth]{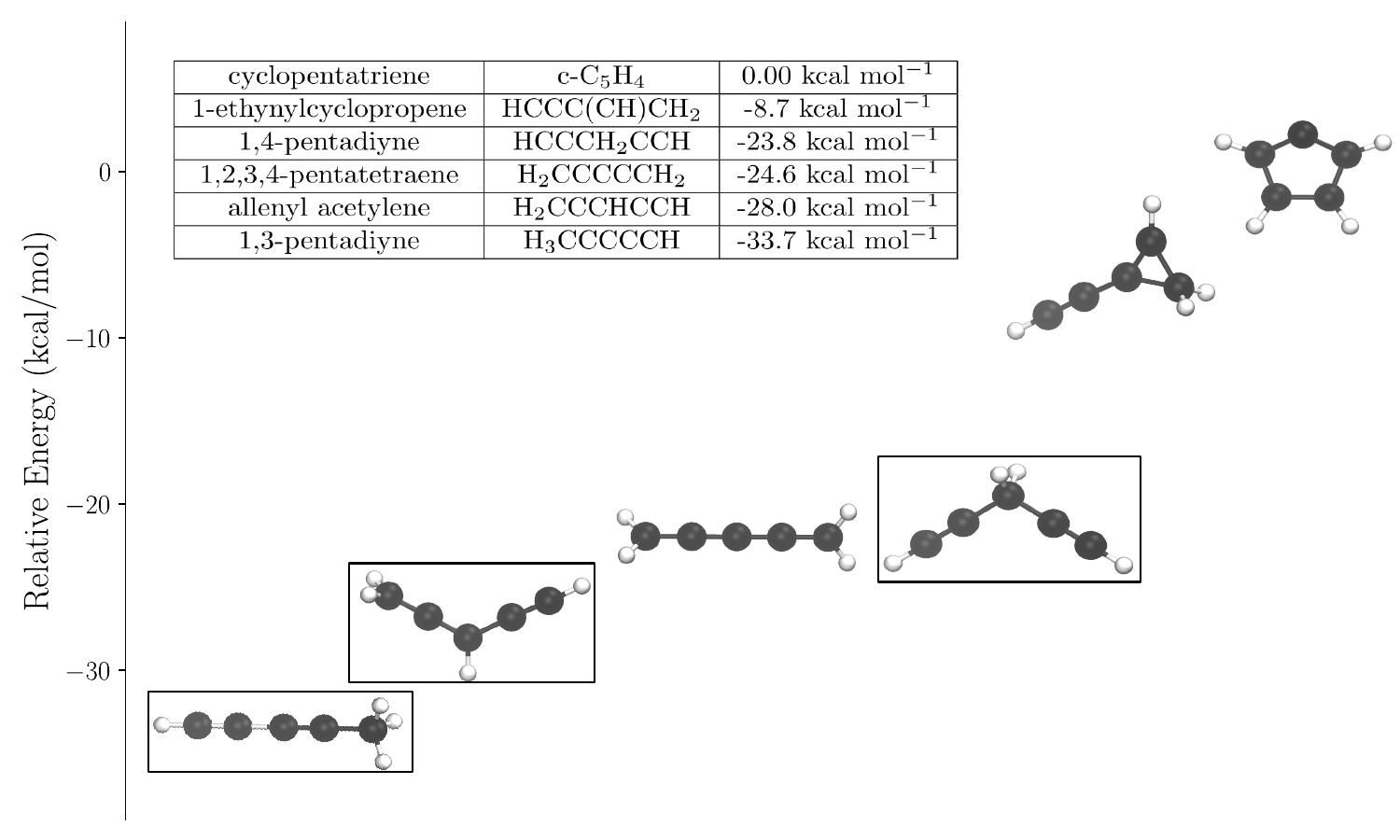}
      \caption{Representation of the relative energy of the five most stable isomers of \ce{C5H4} with respect to the cyclic structure. The energies are calculated at the DLPNO-CCSD(T)/aug-cc-pvtz level with the $\omega$B97XD/def2-TZVPD harmonic zero point energy. The isomers that have been detected in TMC-1 are covered with black boxes \citep{cernicharo2021discovery,fuentetaja2024discovery}.}
         \label{fig:c5h4}
\end{figure}

The different isomeric forms of \ce{C5H4}  motivate the systematic exploration of the energy profile of the \ce{C5H4} molecular formula. The SC-AFIR simulation revealed a total of 454 isomers for this system, where only the most stable ones have been detected with high abundances in the ISM, including  methyl diacetylene (\ce{CH3C4H}), allenyl acetylene (\ce{H2CCCHCCH}) \citep{cernicharo2021discovery} and the 1,4-pentadiyne (\ce{HCCCH2CCH}) \citep{fuentetaja2024discovery} (Figure \ref{fig:c5h4}).

The most stable products recognized during our exploration agree with the five linear isomers of \ce{C5H4} reported in \cite{fuentetaja2024discovery}. Only one cyclic isomer is found, with a relative energy significantly higher than the linear ones, which means that the reaction of \ce{c-C5H5+} with electrons is unlikely to preserve the cyclic structure and the fragmentation into the more stable linear fragments is favored as exemplified in Figure \ref{fig:c5h4}. Interestingly, the products of the dissociative recombination of \ce{c-C5H5+} open a new pathway for the formation of linear carbon chains that have recently been detected in TMC-1. Within the stability ladder below 10 \kcalmol shown in Figure~\ref{fig:c5h4}, only \ce{H2CCCCCH2} has not been directly observed in TMC-1, as it lacks a permanent dipole moment. Together with the high abundance of the precursor of the title reaction and its favorable kinetics, this supports a plausible top-down formation route for these species.

The reaction of \ce{c-C5H5+} with abundant radicals in TMC-1 can potentially induce ring expansion, leading to the formation of larger aromatic compounds. Reactions with atomic carbon (\ce{C}) and methylidyne (\ce{CH}) can compete with electron recombination, and we have aditionally tested the reactions with \ce{CH2} and \ce{CH3}, to which one could add \ce{C2H}, \ce{C3H} and any other carbon bearing abundant radical. We conducted a preliminary investigation using our automated path search protocol, while a more detailed analysis of the reaction mechanisms is postponed to future work. This search revealed remarkably rich reactivity, characterized by the identification of several energetically accessible reaction pathways: \footnote{The precise charge (and spin) state of the products in reactions~\ref{ch3}–\ref{c} is non-trivial and will be addressed in subsequent studies. In the equations presented here, the charge distribution within the products is left open. \rev{For astrochemical modelling, all of the possibilities were included.}}

\begin{align}
   \ce{c-C5H5+ + CH3 &-> c-C6H7^{0/+} + H^{0/+}} \label{ch3} \\  
   \ce{c-C5H5+ + CH2 &-> c-C6H6^{0/+} + H^{0/+}} \label{ch2} \\ 
   \ce{c-C5H5+ + CH &-> c-C6H5^{0/+} + H^{0/+}}  \label{ch} \\
   \ce{c-C5H5+ + C &-> c-C6H4^{0/+} + H^{0/+}}  \label{c} 
\end{align}

It should be noted that if reaction \ref{c} is proven to be effective, it would play a major role in the formation of benzyne (\ce{c-C6H4}), whose ortho isomer was detected in TMC-1 by the QUIJOTE survey \citep{benzyne}. Although several neutral-neutral routes were first proposed to account for the observed abundance of \ce{o-C6H4} \citep{zhang2011formation}, its formation mechanism is still under investigation. Therefore, this reaction may very well represent an alternative pathway to benzyne formation, supporting the bottom-up formation of aromatic molecules in space.

\begin{figure}
   \centering
   \includegraphics[width=\linewidth]{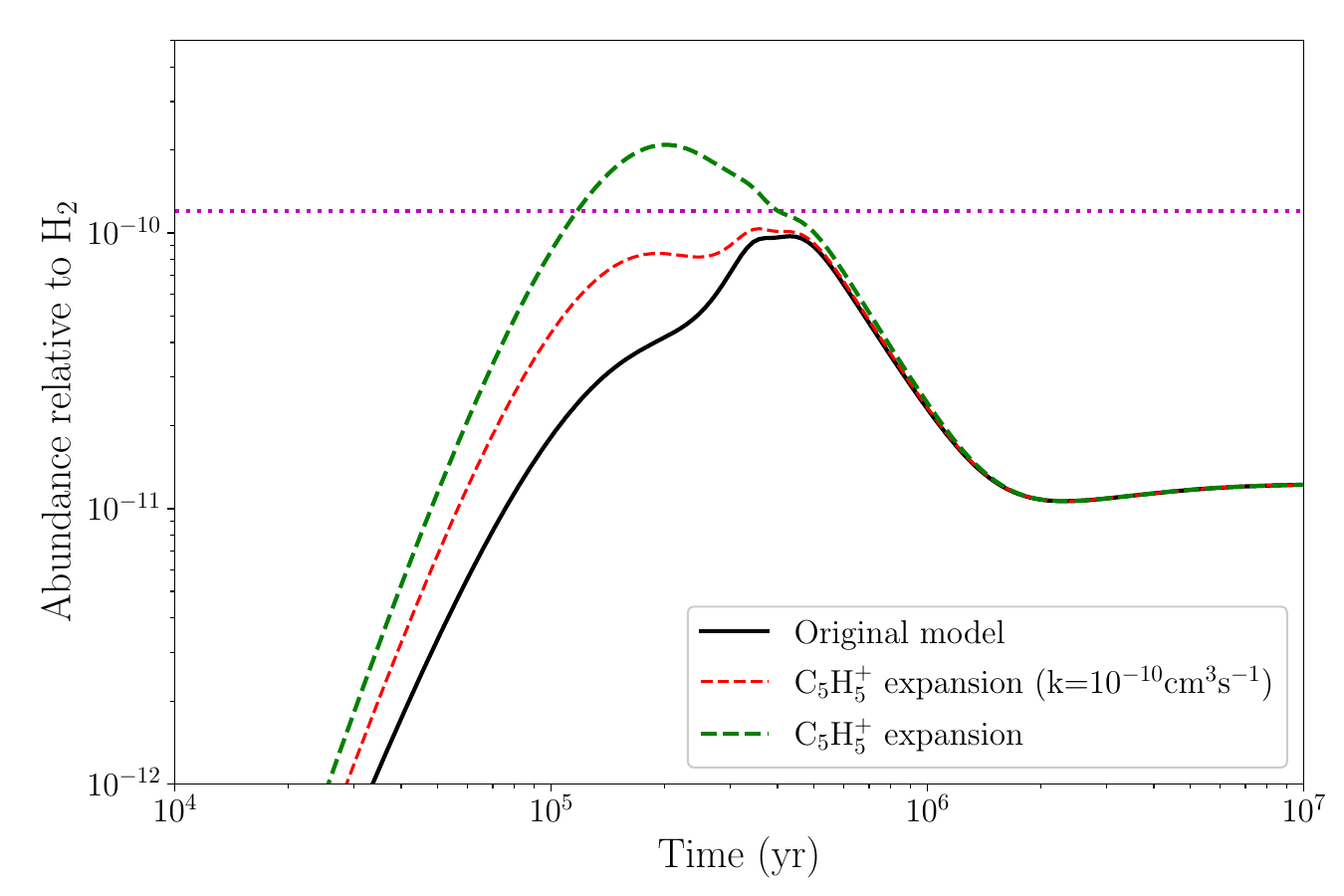}
      \caption{Prediction of the relative abundance of benzonitrile (\ce{C6H5CN}) as a function of time in the chemical model of TMC-1. The black solid line is the abundance predicted by the model before including the ring expansion mechanisms and the dashed lines represents how it is modified considering the formation of the six-membered rings for two possible reaction rate constants, $10^{-9}$ (green) and 10$^{-10}$ (red), see text. The horizontal, dotted line represents the observed abundance of benzonitrile in TMC-1 \citep{cernicharo2021cycles}.}
         \label{fig:c6h5cn}
\end{figure}

The best observationally constrained proxy to assess the impact of reactions \ref{ch3}–\ref{c} is benzonitrile (\ce{C6H5CN}), detected by \citet{mcguire2018detection}. Therefore we incorporated the reactions \ref{ch3}–\ref{c} into the astrochemical model of TMC-1, assuming a rate coefficient of $10^{-9}$ cm$^{3}$ s$^{-1}$ for each, following the same approach as in the exploration of \ce{c-C5H7+} (details in Section \ref{sec:models}). Additionally, we considered rate constants of 10$^{-10}$ cm$^{3}$s$^{-1}$ in our investigation, to evaluate the sensitivity of the model to these reactions. The results, displayed in Figure \ref{fig:c6h5cn}, show that the original model already reproduced reasonably the observed abundance of \ce{C6H5CN}. The inclusion of reactions \ref{ch3}–\ref{c} significantly enhances the abundance of \ce{C6H5CN} below the characteristic timescale of TMC-1 \citep{wakelam2006effect} of around 1$\times$10$^{5}$ years. This time displacement in chemical models is independent of the value of the rate constants and it is in line with some of our latest results, in which the abundance of certain molecules is better reproduced at earlier times \citep{fulvenallene}.

\subsection{The \ce{c-C5H5+ + H2} reaction} \label{sec:c5h5_h2}

Our exploration can be extended to other ion-molecule reactions that were initially considered in Section \ref{sec:models}. The reaction between \ce{c-C5H5+} and \ce{H2} (eq. \ref{eq4}) is predicted to have the strongest impact on the abundance of cyclopentadiene in TMC-1 according to Figure \ref{fig:models}. During the automated search of reaction pathways for the \ce{CH2CCH+ + C2H4} system, we identified the \ce{c-C5H5+ + H2} channel exhibiting an entrance barrier of 3 \kcalmol. However, Figure \ref{fig:fullpes} corresponds to the singlet PES, while the ground state of \ce{c-C5H5+} is a triplet \citep{lee1999study}. Since electronic states play a crucial role in the definition of the reaction pathways, we performed an exploration of the \ce{c-C5H5+ + H2} reaction in the triplet state to confirm its viability and identify the presence or absence of entrance barriers for this scenario.

\begin{figure}
   \centering
   \includegraphics[width=\linewidth]{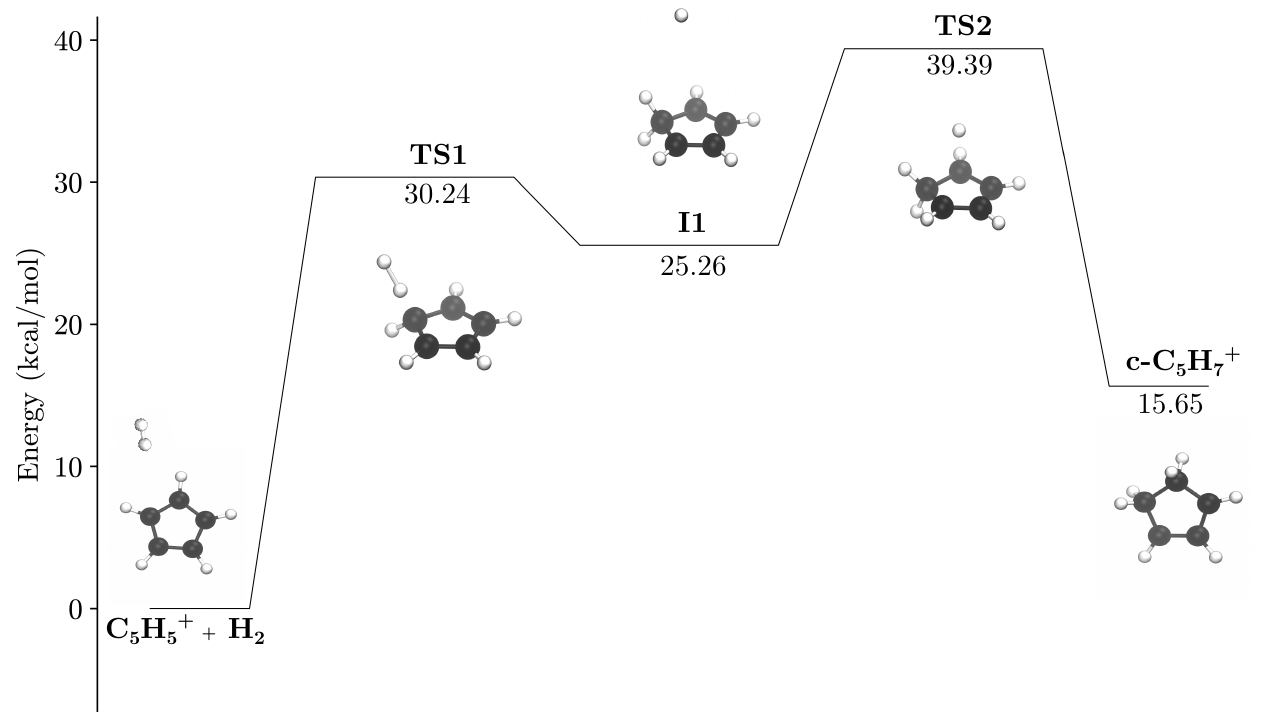}
      \caption{Energy profile of the \ce{c-C5H5+ + H2} reaction in the triplet state. The energies are calculated at the DLPNO-CCSD(T)/aug-cc-pvtz level
      with the $\omega$B97XD/def2-TZVPD harmonic zero point energy.}
         \label{fig:path-triplet}
\end{figure}

Following the methodology employed throughout this project, we performed MC-AFIR calculations to explore the possible association complexes for the \ce{c-C5H5+ + H2} reaction in the triplet state. No evidence of barrierless adduct formation was found, suggesting that the reaction is unlikely to occur under cold interstellar conditions. To validate this result, we constructed the reaction path for \ce{c-C5H5+ + H2}, shown in Figure~\ref{fig:path-triplet}. An entrance barrier of 30.9 \kcalmol was identified, which cannot be overcome in typical interstellar environments. The reaction proceeds through the formation of a post-reactant complex and an additional barrier. In the triplet state, the formation of \ce{c-C5H7+} is endothermic, thereby definitively ruling out \ce{c-C5H5+ + H2} as a viable pathway to \ce{c-C5H7+} in TMC-1. 

\section{Conclusions} \label{sec:astro_implications}

Our results imply a series of far reaching repercussions for the astrochemical community, in particular regarding the reactivity of aromatic hydrocarbons in the ISM. 

As shown in Figure~\ref{fig:models}, the initial model of TMC-1 predicts abundances for \ce{c-C5H6} several orders of magnitude lower than the observed values. Our results indicate that all the ion–neutral routes initially proposed for the synthesis of \ce{c-C5H7+} are not efficient enough to account for the observed abundance of cyclopentadiene in TMC-1, including radiative association of \ce{C2H4 + CH2CCH+}. This has several implications for the bottom-up formation scenario. First, it clearly points to the existence of yet unidentified molecular formation pathways. Second, it may indicate that the efficiencies of the destruction channels are being overestimated. Finally, it could suggest that a fraction of cyclopentadiene is inherited from preceding top-down chemistry. A combination of these factors may well provide the explanation, although we are more inclined to attribute the discrepancy to an incomplete reaction network for this species, whose chemistry remains poorly characterized. As frequently observed in astrochemistry, we cannot discard that the apparent underestimation of cyclopentadiene arises from a similar underestimation of its precursors, particularly the neutral species, as suggested, for example, by \citet{concepcion}.

From Figure \ref{fig:fullpes}, the main reaction pathways can be inferred. The formation of \ce{CH3CCH2+} and \ce{C2H2} from the potential well \textbf{R3}, without going through any intermediate structures, indicates that the reverse reaction, i.e., \ce{CH3CCH2+ + C2H2}, would access the potential energy surface exothermically via the barrierless adduct R3, from which the dominant channel leads to \ce{c-C5H5+ + H2}.  Regarding the \ce{c-C5H5+ + H2} channel in the singlet state to form \textbf{R2}, an activation barrier of 3 \kcalmol is found, demonstrating that the reaction between \ce{c-C5H5+} and \ce{H2} is not barrierless and would not take place under interstellar conditions. Moreover, we it is safe to assume that singlet \ce{c-C5H5+} under interstellar conditions will decay to the triplet state before partaking in any other collision.

We now examine the main product of the title reaction, \ce{c-C5H5+}, which revealed an unexpected role in the formation of detected carbon chains and cyclic molecules alike. In Section \ref{sec:c5h5} we suggest that it can be a leading actor in the complex hydrocarbon chemistry in TMC-1. In addition to our studies, \cite{eyler1984reactivities} demonstrated experimentally that the \ce{C5H5+} cation is reactive toward neutral species such as ethyne (\ce{C2H2}), ethylene (\ce{C2H4}), methylacetylene (\ce{CH3CCH}) and diacetylene (\ce{C4H2}) at room temperatures (300K). None of these reactions are included in the \textsc{UMIST} database, therefore they are not considered in the models to evaluate their impact in colder environments, but using our automated protocol to study them can further enhance our knowledge of the unconventional chemistry protagonized by this radical cation. The reactions of \ce{c-C5H5+} with neutral, abundant species will be further explored in the near future. So far, in the present work, we have proven the formation of the \ce{c-C5H5+} cation with fast rate coefficients at low temperatures, and therefore we propose it to be a key molecule in the chemistry of the interstellar medium, opening the door to new reactivity that was not considered before. Our astrochemical models indicate that the reactions of \ce{c-C5H5+} with small atoms and radicals may play a dominant role in the formation of the best known PAH precursor, benzonitrile, and therefore, benzene and other PAHs in cold dense clouds.

To sum up, we emphasize the importance of the methodological advances achieved in this work. By developing a protocol that combines approximate astrochemical models to identify key reactions with state-of-the-art automated reaction route mapping (\textsc{GRRM23}), quantum-chemical calculations and kinetic modeling, we were able to self-consistently improve the chemical description of TMC-1 through the formation of \ce{c-C5H5+}. This ion has been identified as both a potential precursor to larger aromatic compounds and a possible contributor to the observed isomers of \ce{C5H4} in TMC-1. A major limitation, not addressed in detail here, is the computational cost associated with the automatic reaction discovery protocol, which currently prevents the simultaneous exploration of tens or hundreds of reactions. For this reason, astrochemical models are invaluable for rapidly filtering promising reactions, but they remain insufficient on their own. The development of broadly applicable machine-learned potentials for astrochemically relevant systems could greatly accelerate the exploration of potential energy surfaces, and therefore the entire protocol. Achieving this objective remains a long-term goal for our group.

\begin{acknowledgements}
\rev{The authors want to thank the reviewer of this manuscript, Emilio Martinez Nuñez, for his thorough revision of our work.} G.M. acknowledges the support of the grant RYC2022-035442-I funded by MCIU/AEI/10.13039/501100011033 and ESF+. G.M. also received support from project 20245AT016 (Proyectos Intramurales CSIC). We acknowledge the computational resources provided by the DRAGO computer cluster managed by SGAI-CSIC, and the Galician Supercomputing Center (CESGA). The supercomputer FinisTerrae III and its permanent data storage system have been funded by the Spanish Ministry of Science and Innovation, the Galician Government and the European Regional Development Fund (ERDF). M.A., C.C. and J.C. acknowledge funding support from the Ministerio de Ciencia, Innovación y Universidades through project PID2023-147545NB-I00, M.M., G.M. and O.R. through the project PID2024-156686NB-I00 and O.R through the project PID2021-122549NB-C21. 

\end{acknowledgements}

%
%

\bibliographystyle{aa}
\bibliography{ref}

\begin{appendix}
\section{Derivation of the radiative association rate constant for the \ce{C5H7+} channel} \label{sec:appA_ratiative_association}

In the main text we conclude that the radiative association channel for the \ce{CH2CCH+ + C2H4} reaction, leading to \ce{C5H7+} is negligible in comparison with the bimolecular channels. Determining the radiative association rate constant ($k_{\rm rad}$) in a PES landscape as complex as the one shown in Figure \ref{fig:fullpes} is a daunting task. Therefore, in this appendix we derive a simplified value for such rate constant that is the one that we use for comparison in the main text and we report in Table \ref{tab:fit}. Considering that there are open bimolecular channels in the reaction, we can take the back dissociation rate constant $k_{-1}$ out of equation \ref{eq:rad_total} of the main text:

\begin{figure}
   \centering
   \includegraphics[width=\linewidth]{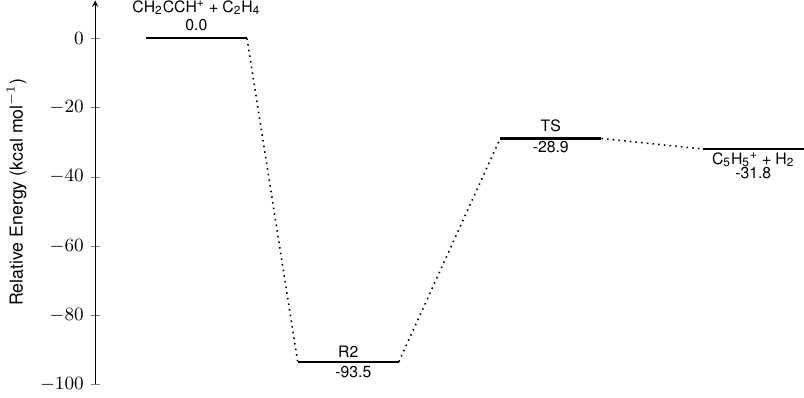}
      \caption{Reduced potential energy surface (from Figure \ref{fig:fullpes} used in the model of radiative association.)}
         \label{fig:pes_reduced_radiative}
\end{figure}

\begin{equation} \label{eq:rad_partial}
   k_{\rm rad} = \frac{k_{1} \, k_{r}}{k_{2} + k_{r}},
\end{equation}
where all previously defined quantities are retained. In our approximate scheme to determine $k_{\rm rad}$, a few simplifying assumptions were introduced. First, although the association complex resulting from the \ce{CH2CCH+ + C2H4} reaction corresponds to \textbf{R1} (Figure~\ref{fig:r1}), we assume for the purpose of our derivation that the complex is captured in \textbf{R2} with the same capture rate constant ($k_{1}$). This is justified because \textbf{R2} represents the deepest well on the sampled PES and the configuration from which radiative emission would most likely yield the product of interest, \ce{c-C5H7+}. Second, we assume that k$_{2}$, which in principle accounts for all possible bimolecular transitions out of the PES leads exclusively to the formation of \ce{c-C5H5+ + H2}, without involving any pre-reactive complexes (PRCs), in contrast with the full scheme shown in Figure~\ref{fig:fullpes}. Within this framework, we effectively decouple the radiative association probability from potential isomerizations across the entire PES, thereby reducing the ``effective'' surface to that depicted in Figure~\ref{fig:pes_reduced_radiative}. These approximations are valid only when $k_{\rm rad}$ is several orders of magnitude smaller than the competing channels, which, as will be shown below, is indeed the case. In Equation~\ref{eq:rad_partial}, $k_{1}$ and $k_{2}$ are obtained from our master equation analysis: $k_{1}$ represents the capture rate constant, while $k_{2}$ corresponds to the (microcanonical) forward rate constant leading to \ce{c-C5H5 + H2} (i.e., the PRC). Therefore, in order to apply Equation~\ref{eq:rad_partial}, it only remains to determine $k_{r}$.

\begin{figure}
   \centering
   \includegraphics[width=\linewidth]{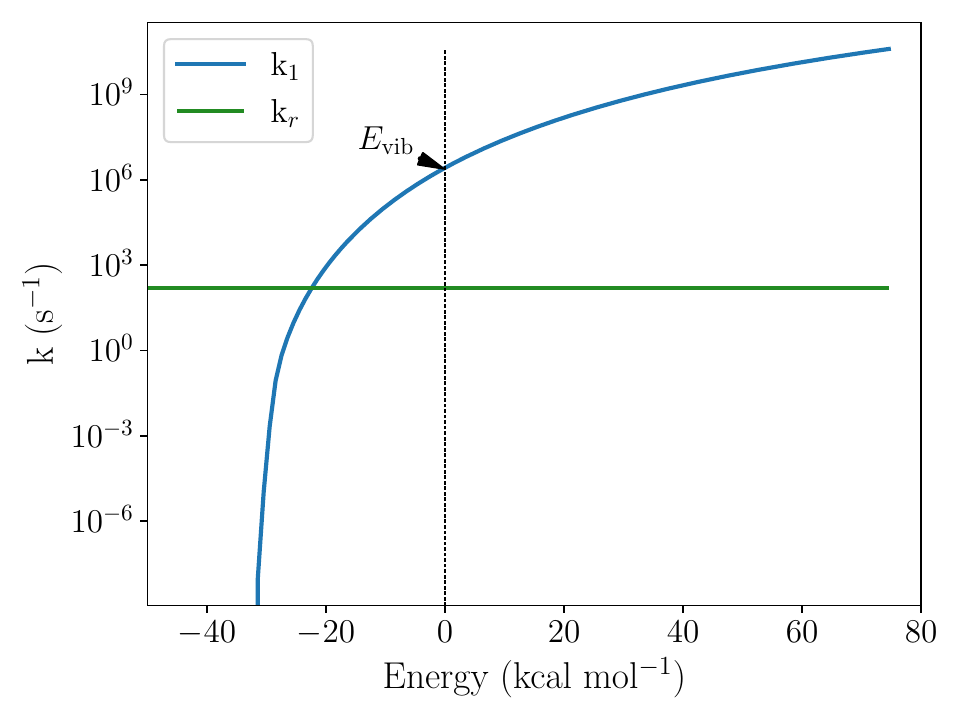}
      \caption{Radiative and unimolecular forward rate constants for the competing processes of the formation of \ce{c-C5H7+} or \ce{c-C5H5+ + H2}, the energy of the activated complex E$_{\rm vib}$ is taken as the origin zero energy. Energies below 0 \kcalmol are not physically feasible for this system.}
         \label{fig:k_r_reduced_radiative}
\end{figure}

Following \citet{herbst_approach_1982} and more recently \cite{tennis_radiative_2021}, the rate constant for the radiative emission rate, assuming that the activated complex can be estabilized by the emission of a single photon, assuming harmonic approximation and statistical distribution of the association energy between the reactants is:

\begin{equation} \label{eq:kr}
   k_{r} = \dfrac{E_{\rm vib}}{s}\sum_{i=1}^{s}\dfrac{A^{(i)}_{1 \rightarrow 0 }}{h\nu_{i}}.
\end{equation}
In the above equation $E_{\rm vib}$ is the vibrational energy of the activated complex (\textbf{R2} in this case), which is the well energy (93.5 \kcalmol with the addition of any additional collision energy), $s$ is the number of molecular vibrations partaking in the ergodic energy distribution, $A^{(i)}_{1 \rightarrow 0 }$ are the Einstein spontaneous emission coefficients for each vibration ($i$), $h$ is the Planck constant and $\nu_{i}$ the fundamental frequency of mode $i$. The $A^{(i)}_{1 \rightarrow 0 }$ can be determined as:
\begin{equation}
   A^{(i)}_{1 \rightarrow 0 } = \dfrac{8\pi}{c}\nu_{i}^{2}I_{i}
\end{equation}
where, in addition to the speed of light constant $c$, the only remaining quantity to define is $I_{i}$ that is the infrared intensity of each normal mode. The results for $k_{r}$ can be visualized in Figure \ref{fig:k_r_reduced_radiative}. From the graph, it is evident that at $E_{\rm vib}$, k$_{1}$ clearly dominates, which represents the optimistic case for radiative association where the relative energy for the collision is zero, and the available energy to the system is only the association energy at \textbf{R2}. Plugging the obtained numbers into Equation \ref{eq:rad_partial} we obtain a radiative association rate constant of $k_{\rm rad}$$\leq$3.25$\times$10$^{-14}$ cm$^{3}$ s$^{-1}$, which remain rather constant at low temperatures, but that should decay even further when $k_{1}$ rises at higher temperature. \rev{The $k_{\rm rad}$ is around 4-5 orders of magnitude lower than the competing process.} This is the value of the rate constant reported in Table \ref{tab:fit} and used to establish comparisons in the main text. To conclude, we also tested the validity of the method shown in \citep{tennis_radiative_2021} with a more accurate representation of the radiative transfer process, proposed in \cite{cernicharo2023magnesium}. The value of k$_{r}$ according to \cite{cernicharo2023magnesium} is determined to be 3.86 $\times$10$^{2}$ s$^{-1}$, compared to 1.53 $\times$10$^{2}$ s$^{-1}$ obtained from Equation \ref{eq:kr}. No significant differences were observed between the models in terms of the radiative emission rate constant, that is parameter that would change in equation \ref{eq:rad_partial}, which supports the reliability of our results using a simplified model.

\end{appendix}
\end{document}